\documentclass[10pt,twocolumn,english,aps,pra,longbibliography,superscriptaddress,amsmath,amssymb,floatfix]{revtex4-2}

\usepackage{graphicx, xcolor, nicefrac} 
\usepackage{lipsum}

\definecolor{dgreen}{RGB}{0,102,51}
\definecolor{orangered}{RGB}{255,69,0}
\usepackage[colorlinks=true,urlcolor=blue,citecolor=blue,linkcolor=blue]{hyperref}
\usepackage{amsmath} 
\usepackage{physics}
\usepackage[english]{babel}
\usepackage[letterpaper,twocolumn,top=2cm,bottom=2cm,left=1.5cm,right=1.5cm,marginparwidth=.75cm]{geometry}

\usepackage{braket} 

\makeatletter
\renewcommand*\env@matrix[1][\arraystretch]{%
  \edef\arraystretch{#1}%
  \hskip -\arraycolsep
  \let\@ifnextchar\new@ifnextchar
  \array{*\c@MaxMatrixCols c} 
}
\makeatother

\date{}

\begin{document}
\setlength{\parskip}{0pt}

\title{Local Rydberg blockade regimes for disk graph embedding and quantum optimization}
\author{Elie Bermot}
\email{elie.bermot@pasqal.com}
\affiliation{PASQAL, 24 Av. Emile Baudot, 91120 Palaiseau, France}
\affiliation{Université Paris Cit\'e, CNRS, IRIF, Paris, France}
\author{Lucia Valor}
\affiliation{PASQAL, 24 Av. Emile Baudot, 91120 Palaiseau, France}
\author{Wesley Coelho}
\affiliation{PASQAL, 24 Av. Emile Baudot, 91120 Palaiseau, France}
\author{Louis-Paul Henry}
\affiliation{PASQAL, 24 Av. Emile Baudot, 91120 Palaiseau, France}
\author{Natalie Pearson}
\affiliation{PASQAL, 24 Av. Emile Baudot, 91120 Palaiseau, France}
\date{\today}
\begin{abstract}

Rydberg atom arrays are a powerful platform for solving combinatorial optimization problems, owing to the Rydberg blockade mechanism, which imposes effective constraints on simultaneous atomic excitations. These constraints have enabled the encoding of the Maximum Independent Set (MIS) problem on unit disk graphs, where atoms interact within a fixed, globally defined blockade radius. However, this restriction limits the class of addressable problems. A natural extension is to consider disk graphs, which generalize unit disk graphs by allowing arbitrary disk radii and correspond to the intersection graphs of disks in the plane. Embedding such graphs in Rydberg systems requires moving beyond the standard, globally uniform blockade model. In this work, we introduce a local Rydberg blockade regime, which emerges when local drives are applied to different pairs of atoms involved in a potential interaction. We develop a general theoretical framework for this regime and propose two novel metrics, the correlation matrix and the maximum independence violation, 
to quantify the quality of the embedding.  Using these metrics, we demonstrate that disk graphs can be meaningfully embedded into Rydberg atom arrays under local drive schemes, thereby expanding the landscape of quantum-addressable optimization problems. Finally, when evaluating approximate solutions of the MIS problem, characterized by near-optimal independent sets, local drive approaches exhibit significantly improved performance over global ones. These results highlight the practical advantage of local blockade engineering for approximate combinatorial optimization and open a path toward leveraging the analog capabilities of Rydberg platforms beyond conventional geometric constraints.

\end{abstract}
\maketitle

\section{Introduction}
\label{sec:intro}

Neutral-atom quantum processors have become a compelling platform for solving combinatorial optimization problems, owing to their ability to implement both analog and digital quantum algorithms~\cite{saffman2010quantum, white_paper, adams2019rydberg}. A central mechanism enabling such computations is the Rydberg blockade: when an atom is excited to a Rydberg state, its strong dipole-dipole interaction shifts the energy levels of neighboring atoms, thereby preventing their simultaneous excitation. This interaction has been extensively studied and validated experimentally~\cite{PhysRevLett.87.037901, Urban_2009, Wilk_2010, Guttridge_2023}, and forms the basis of entangling gates in digital quantum computing~\cite{saffman2020symmetric, levine2019parallel, pelegri2022high} as well as constrained dynamics in analog settings~\cite{NguyenPichler2023, scholl2021quantum, weimer2010rydberg, morgado2021quantum}.

One particularly promising application of the Rydberg blockade is the encoding and solving of the Maximum Independent Set (MIS) problem. By mapping each atom to a graph node and forbidding simultaneous excitation of atoms within a given blockade radius, the system naturally evolves within the subspace of independent sets, i.e., sets in which no two adjacent vertices are simultaneously occupied. The MIS problem, which seeks the largest such set, is known to be NP-hard~\cite{Garey1978S}, and has therefore attracted interest as a candidate for demonstrating quantum advantage. This approach has been implemented on unit disk graphs, where vertices are placed on a plane and connected if their Euclidean distance falls below a fixed threshold~\cite{Ebadi_2022, Dalyac_2021, da_Silva_Coelho_2023, dalyac2024graphalgorithmsneutralatom}. In principle, such graphs align naturally with the uniform blockade constraint imposed by global laser drives and constant interatomic distances.

However, the class of embeddable graphs within this framework is fundamentally limited. Unit disk graphs form a geometrically simple subclass of intersection graphs, and when this is further restricted by hardware limitations such as finite optical resolution, minimum atom spacing, and lattice connectivity, the result is a very narrow family of problem instances that can be realized experimentally. Moreover, many of the unit disk MIS instances studied to date are either structurally regular or classically solvable faster than current quantum approaches~\cite{cazals2025identifyinghardnativeinstances}, raising important questions about the broader computational reach of existing methods. A natural direction for exploring the potential of neutral-atom platforms is thus to expand the set of embeddable graphs toward families that support richer and harder instances, while still leveraging the analog structure of the Rydberg system.
At the same time, it has become increasingly clear that analog quantum protocols should be evaluated not only by their ability to reach exact ground states, but also by their capacity to produce high-quality approximate solutions~\cite{Munoz_Bauza_2025}. In the context of the MIS problem, this corresponds to identifying configurations close to the optimal independent set which captures the practical regime accessible to noisy, finite-time dynamics.

To this end, we investigate the embedding of disk graphs, a natural generalization of unit disk graphs in which each vertex corresponds to a disk of arbitrary radius in the plane, and edges are present when two disks intersect. Disk graphs can represent more complex, heterogeneous structures and are known to encode classically challenging optimization problems~\cite{van2006representation}. Crucially, embedding such graphs requires moving beyond the standard, idealized picture of the Rydberg blockade based on a global drive and a single fixed blockade radius. In practice, the blockade effect is a soft, continuous phenomenon, sensitive not only to interatomic distance but also to the amplitude and detuning of the laser drive, especially in the presence of inhomogeneous driving conditions. As such, a more general theory is needed to describe how the Rydberg blockade operates under local driving, where each atom may experience a distinct laser amplitude.

In this work, we introduce and characterize the local Rydberg blockade regime, a setting in which the driving field is no longer homogeneous. We develop a theoretical framework to describe this regime and show how it gives rise to amplitude-dependent generalizations of the blockade radius. Building on this foundation, we propose two complementary metrics to evaluate how well a given spatial and amplitude configuration of atoms realizes the embedding of a target disk graph. The first, the correlation matrix, captures joint excitation probabilities and reveals the effective adjacency structure induced by the Rydberg interactions. The second, the maximum independence violation, provides a quantitative measure of how well the embedding respects the independent set properties of the desired graph. Together, these tools allow us to assess the fidelity of graph encodings in both global and local driving regimes.

Using this framework, we show that disk graphs, which cannot be faithfully embedded under global driving, become accessible under local drive schemes. We demonstrate that local addressing substantially enhances approximate optimization performance relative to global driving for instances that lie outside the unit disk class. Our results thereby extend the capabilities of neutral-atom quantum processors and make them applicable to a broader class of geometrically and computationally complex optimization problems.

The paper is organized as follows. In Section~\ref{sec:lucia}, we introduce the local Rydberg blockade regime for two-atom systems and rederive the interaction structure for both sequential and global driving. We analyze how the resulting effective blockade radius depends on the driving configuration and detuning. In Section~\ref{sec:natalie}, we extend this analysis to many-body systems and introduce the correlation matrix and maximum independence violation as tools to evaluate the faithfulness of graph embeddings under local blockade conditions. In Section~\ref{sec:elie}, we apply this framework to solving the Maximum Independent Set problem on disk graphs and show that local driving enhances approximate optimization performance relative to global approaches. We conclude by discussing the implications of our findings for scaling neutral-atom systems and tackling more computationally demanding problems in the analog regime.

\section{Local Rydberg blockade}\label{sec:lucia}
The Rydberg blockade mechanism is typically introduced as a binary constraint: two atoms within a cutoff distance cannot be simultaneously excited. However, this sharp cutoff is only an approximation. In practice, the blockade effect is a continuous phenomenon that depends on interatomic distance, detuning, and driving amplitude. In particular, the impact of local amplitudes, where atoms are driven with different strengths, is not captured by the standard global blockade model.

The goal of this section is to quantitatively re-express and characterize the Rydberg blockade in a more general setting. We first focus on what we refer to as the local blockade regime, in which atoms experience distinct local amplitudes. We then analyze how the blockade behavior varies across different drive configurations and derive consistent, amplitude-dependent definitions of blockade strength and blockade radius. This lays the foundation for the many-body analysis and graph-embedding applications presented in the following sections.

The Rydberg blockade mechanism~\cite{Urban_2009, gaetan2009observation} can be understood through an energy-level representation in the 
$z$-basis, where the basis states are the ground state $\ket{g}=\ket{0}$, and a Rydberg state, $\ket{R}=\ket{1}$. In this configuration, the Hamiltonian of a neutral atom system is equivalent to that of a Transverse Field Ising Model (TFIM). With $\hbar = 1$, this takes the form 
\begin{equation}\label{eq:ryd_ham}
    \mathcal{H} = \dfrac{1}{2}\sum_i \Omega_i \sigma_i^x -\sum_i \delta_i n_i + \sum_{i<j} \dfrac{C_6}{r_{ij}^6}n_in_j,
\end{equation}
where $r_{ij}$ is the distance between atoms $i$ and $j$, $n_i=(1-\sigma_i^z)/2=\ketbra{R}{R}$ represents the Rydberg state occupancy of atom $i$, and $\sigma_i^x$ is the corresponding Pauli X operator. The parameters $\Omega_i(t)$ and $\delta_i(t)$ correspond to the amplitude and detuning of the laser driving the transition between $\ket{g}$ and $\ket{R}$. The interaction term, characterized by the van der Waals coefficient $C_6$, describes the long-range interaction between Rydberg atoms, which scales as $1/r^6$ and depends on the specific Rydberg level targeted. For instance, for a system using the Rydberg state with principle quantum number $n = 70$, $C_6$ is $862 \, \text{GHz} \cdot \mu\text{m}^6$, and for $n = 82$, $C_6$ is $5559 \, \text{GHz} \cdot \mu\text{m}^6$~\cite{_ibali__2017}. In this project, we use the Rydberg state with principal quantum number $n=70$.

As seen in the last term of Eq.~\eqref{eq:ryd_ham}, for a pair of atoms, the van der Waals interaction shifts the energy of the doubly excited state $\ket{RR}$ by a factor of $C_6/r_{ij}^6$. Assume the atoms are driven by a laser resonant with the $\ket{g} \leftrightarrow \ket{R}$ transition for both atoms ($\delta = 0$). If the atoms are separated by a distance $r \ll r_B = (C_6/\Omega)^{1/6}$, the van der Waals interaction ($C_6/r^6\gg \Omega$) shifts the energy of the double excited state $\ket{RR}$ enough, so that it is no longer resonant with the driving laser, and the driving will result in transfer to the entangled state $(\ket{gR} + \ket{Rg})/\sqrt{2}$.

From this, one can intuitively infer that two atoms are ``blockaded'' when the dominant energy scale is the interaction shifting $\ket{RR}$, i.e. $\Omega+\delta\ll C_6/r_{ij}^6$, and conversely ``non-blockaded'' when $\Omega+\delta\gg C_6/r_{ij}^6$. This understanding is used to deduce the distance between atoms $i$ and $j$ within which only one of these atoms may be excited. This distance is generally termed as the ``blockade radius''~\cite{PhysRevLett.85.2208, PhysRevLett.87.037901} and is usually written as
\begin{equation}\label{eq:rb_sequential}
    r_B = \left(\dfrac{C_6}{\Omega+\delta}\right)^{1/6}.
\end{equation}

\begin{figure*}\includegraphics[width=0.95\textwidth]{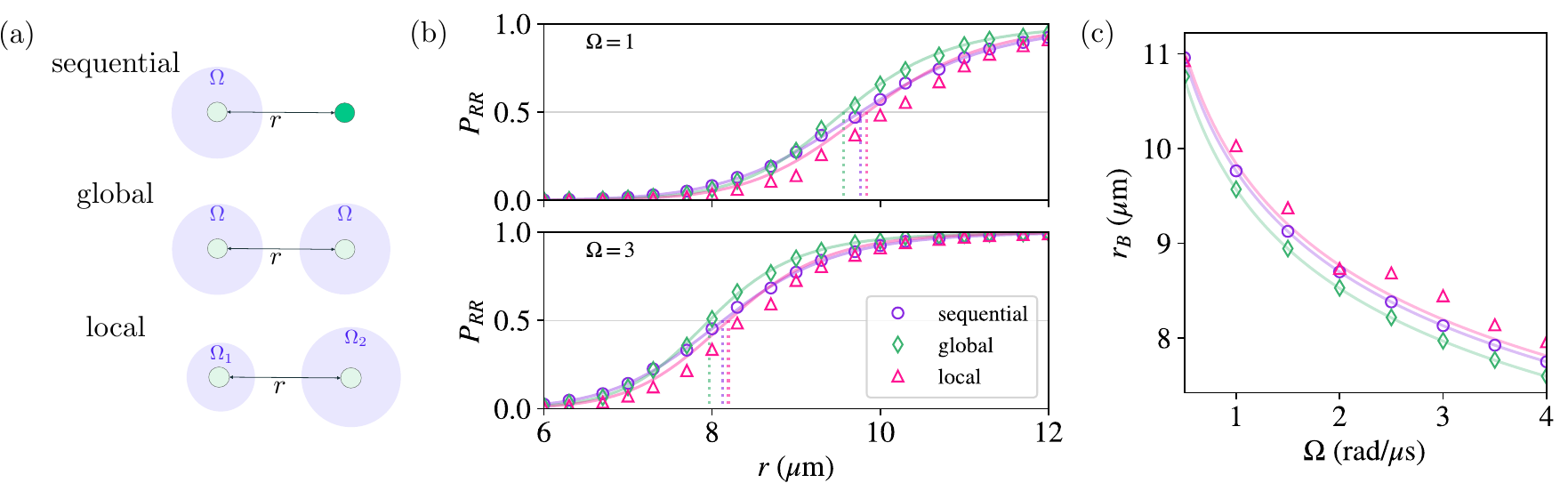}
  \caption{(a) The three driving scenarios are illustrated, where the light (dark) green circles correspond to atoms initialized in the ground (excited) state and driven with lasers with amplitudes $\Omega$. (b) The maximum probability of measuring both atoms in the Rydberg state during a quench with average amplitudes 1 rad/$\mu$s (upper) and 3 rad/$\mu$s (lower) for each driving scenario. Markers indicate simulated results over 50$\mu$s, solid lines are calculated using the equations mentioned in the main text. Dotted lines show where the calculated values cross $P_{RR}=0.5$ and hence define the blockade radius $r_B$. (c) The simulated (markers) and calculated (solid lines) values of $r_B$ for a range of amplitudes. For the locally driven case the average amplitude is $\Omega$ and the ratio is $\Omega_2 / \Omega_1 = 3$ }\label{fig:lucia}
\end{figure*}

To further characterize the crossover between the blockaded and non-blockaded regimes, we introduce a quantitative measure of blockade strength. A natural choice is the maximum occupation of the doubly excited state over time, defined as $P_{RR} = \max\limits_{t} \left| \braket{\psi(t)|RR}\right|^2\in [0,1]$. For a system driven over a sufficiently long duration, we expect $P_{RR}$ to approach a value of 1 if the atoms are very far apart and the system is free to explore all energetically and symmetrically permitted states, including $\ket{RR}$. If, on the other hand, the atoms are blockaded to some extent, we expect the value of $P_{RR}$ to be bounded above by a quantity related to the extent of the blockade. As such, we quantify blockade as $1-P_{RR}$. With this metric defined we can concretely define the blockade radius to be the distance at which $P_{RR}=0.5$.

Figure~\ref{fig:lucia}(a) depicts three different driving scenarios that we investigate, by evaluating the doubly excited state probability $P_{RR}$ as a function of the interatomic separation $r \equiv r_{12}$ and for various amplitudes, as shown in Figure~\ref{fig:lucia}(b). We consider three systems of increasing complexity, each providing insight into a different aspect of the Rydberg blockade. The first is a sequentially driven system (labeled S, with associated blockade radius $r_B^S$ and probability $P_{RR}^S$), where one atom is excited to $\ket{R}$ before the second is driven with a constant local amplitude. This case corresponds to the conventional intuition behind the Rydberg blockade, where the excitation of one atom shifts the energy levels of nearby atoms out of resonance. It is analytically tractable and we find 
\begin{equation} \label{eq:prr_pipulse}
    P_{RR}^{S}(r) = (1+\left[\Omega^{-1}\left(C_6/r_{ij}^6-\delta\right)\right]^2)^{-1},
\end{equation}
resulting in a blockade radius, $r_B^S$ identical to that of the commonly used expression for the blockade radius of Eq.~\eqref{eq:rb_sequential}. 

The second scenario is a simultaneous global drive (G, with $r_B^G$ and $P_{RR}^G$), where both atoms start in $\ket{g}$ and are driven simultaneously with the same amplitude. This setup reflects the standard operating regime of many-body analog Rydberg experiments, but is more complex to analyze and leads to modified blockade behavior compared to the sequential case. More specifically, this can be computationally solved with a specialized software, such as Wolfram Mathematica \cite{Mathematica}, resulting in a higher $P_{RR}^{G}$ for the same values of $r$ and $\Omega$ as shown in Fig. \ref{fig:lucia}(b), and therefore resulting in a lower blockade radius $r_B^G$. The full equation of $P_{RR}^G$ is printed in Appendix \ref{app:lucia}, Eq.~\eqref{eq:prr_globpulse}. Curiously, we observe that a very good numerical approximation of the resulting blockade radius is $r_B^G = 0.98 r_B^S$, shown in Fig. \ref{fig:lucia}(c).

Finally, the simultaneous local drive case (L, with $r_B^L$ and $P_{RR}^L$) extends this to the setting where atoms are driven simultaneously but with different local amplitudes, representing the local blockade regime introduced in this work. This case captures amplitude asymmetry in a minimal two-atom system and serves as the foundation for generalizing to inhomogeneous many-body dynamics. In this case, we drive the system with the same average value of the amplitude as in the former driving scenarios ($\Omega = (\Omega_1 + \Omega_2)/2 \in\{1, 3\} $ rad/$\mu$s), but where the ratio between them is $\Omega_1/\Omega_2 = 3$. In this case, we cannot solve the system exactly, but we can find a good approximate form for $P_{RR}^L$ numerically using a method presented in Appendix~\ref{app:local_rydberg}. For this locally driven system, we observe that the corresponding blockade radius $r_B^L$ for this set of amplitudes $\Omega_i$ satisfies

\begin{equation}
    r_B^L(\Omega_1, \Omega_2) \sim 0.49 \left[ r_B^G(\Omega_1) + r_B^G(\Omega_2) \right].
\end{equation}

The simulated and calculated values for $r_B$ are shown in Figure \ref{fig:lucia}(c). Taken together, this provides us with two outcomes. The first is a better way to quantify blockade--this can be useful when embedding graphs in neutral atom quantum systems. The second is a rigorous definition of the blockade radius and its pairwise behavior, most notably in the presence of local drive. In the next sections, we will investigate the implications of using these developments for graph embedding in many body systems.

\section{Embedding Disk Graphs}\label{sec:natalie}

Having developed a deeper understanding of the Rydberg blockade phenomenon, particularly in the presence of local amplitudes, we now apply this framework to the problem of graph embedding using the Hamiltonian of a Rydberg atom Quantum Processing Unit (QPU). Specifically, we demonstrate how the generalized blockade model enables the embedding of a broader class of graphs, namely disk graphs~\cite{breu1998unit}, by appropriately selecting local amplitudes and atom positions. A disk graph is a type of geometric graph where each vertex represents the center of a disk in the Euclidean plane, and edges are drawn between vertices if and only if the corresponding disks overlap. If all radii are equal, this reduces to a unit disk graph. The key insight from Section~\ref{sec:lucia} is that local driving induces local blockade radii, so each atom has its own radius and each pair of atoms interacts under a distinct blockade condition.. This motivates our approach: by assigning different local amplitudes, we can realize disk graph connectivity. We propose a deterministic method to embed disk graphs using local blockade radii and assess its performance using our previously defined metrics, namely the correlation matrix and the maximum independence violation probability. These embeddings serve as the basis for solving combinatorial optimization problems, such as MIS, in the next section.

\begin{figure*}[t!]\includegraphics[width=0.95\textwidth]{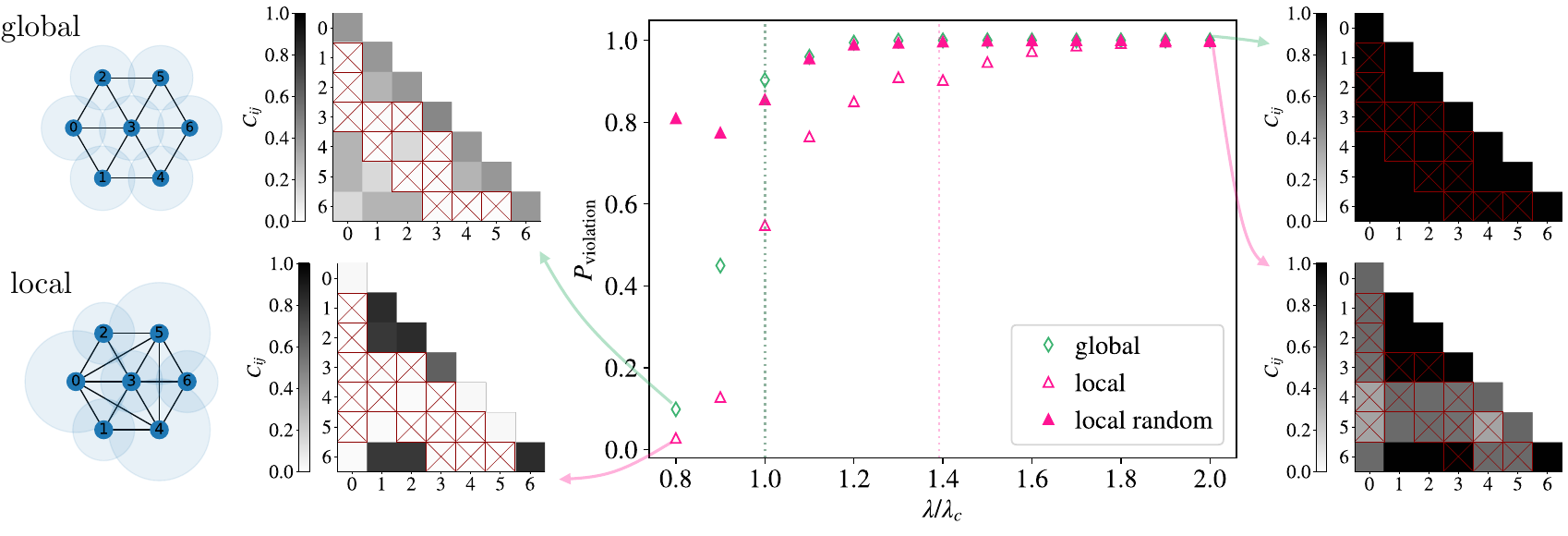}
\caption{Graph embeddings for unit disk graphs (top) and disk graphs (bottom) are evaluated using correlation matrices and maximum independence violation metrics. Both graphs have the same atom position, but local radii result in different connectivity. The radii, $r_{B, i}=0.98\sqrt[6]{C_6/\Omega_i}$, are implemented using a global drive of $\Omega_i=\pi$ for the unit disk graph (green open diamonds)  or a local drive with $\Omega_i=\pi/20$ for $i\in (0, 4, 5)$ for the the disk graph (pink open triangles). A local drive implementation using the same set of local amplitudes but applied to random atoms (pink filled triangles) is shown as a control. A 100$\mu$s quench is performed with the atomic register scaled by $\lambda$, where $\lambda=\lambda_c$ (black dotted line) is the point at which the first edge in the graph is expected to break. Full disconnection of the unit disk graph also occurs at this point, whereas full disconnection of the disk graph is shown by the pink dotted line. The maximum correlation matrix is measured at connected ($\lambda/\lambda_c=0.8$) and disconnected ($\lambda/\lambda_c=2$) scaling values, and overlapped by red crosses which indicate edges in the graph.
}
\label{fig:natalie}
\end{figure*}

To test this disk graph embedding strategy, we perform a quench $\ket{\psi(T)}$ of duration $T = 100 \mu s$ using the QPU Hamiltonian of Eq.~\eqref{eq:ryd_ham},  under two configurations. The first is a global pulse with amplitude $\Omega $, corresponding to the Rydberg blockade radius for the unit disk graph; the second is a scheme of local amplitudes $\Omega_i$, corresponding to the local Rydberg blockade radii. The resulting final states encode the target graph structures according to the blockade behavior described in previous sections and shown in Figure \ref{fig:natalie}.
 
To evaluate the quality of the embedding, we characterize the graphs based on their independent set properties. Let us recall that, for a given graph $G = (V, E)$, an independent set is a subset $S \subseteq V$ such that no two vertices in $S$ are connected, i.e., $\forall u, v \in S, (u, v) \notin E$. We define the \textit{maximum independence violation probability}, $P_{\text{violation}}$, as the maximum value of the expectation value of the projector onto the subspace of non-independent sets over all $t$ during the quench,

\begin{equation}
    P_{\text{violation}} = \max_{t} \braket{\psi(t) | \Pi_{\text{non-independent}} | \psi(t)},
\end{equation}

where $\Pi_{\text{non-independent}}$ projects onto states that do not form an independent set, exciting at least one pair of atoms representing connected nodes. This metric reduces to $P_{RR}$ for the two atom case, assuming the two atoms share an edge. We repeat this for different distance scales by scaling the atomic coordinates by a parameter $\lambda$. By assuming the same underlying graph, we expect that, for some $\lambda$, atoms will be too distant for their interaction to well implement an edge between them. We refer to this as the `breaking' of an edge, and define a critical value of the scaling $\lambda_c$ to be the value at which the first edge is broken. The behavior of $\lambda_c$ differs between unit disk graphs and disk graphs. For unit disk graphs, $\lambda_c$ corresponds to the point where the maximum separation between connected atoms exceeds the blockade radius. In contrast, for disk graphs, edges can break between atom pairs that are not necessarily the furthest apart, since blockade radii vary. For $\lambda < \lambda_c$, the maximum independence violation should be minimal, as all atoms remain blockaded. However, for $\lambda > \lambda_c$, at least one expected edge in the embedded graph may be missing. Therefore, we expect the mean independence violation probability to exhibit a behavior similar to $P_{RR}$, as described in the previous section.

This encoding is further studied by looking at the maximal correlation matrix, defined as

\begin{equation}
    C_{ij} = \max_{t}\braket{\psi(t) | n_i n_j | \psi(t)} .
\end{equation}

A small value of $C_{ij}$ implies that atoms $i$ and $j$ cannot be excited at the same time, while a large value implies that both atoms can be. Therefore, if the maximum independence violation is small, it is expected to also have a small value of the maximal correlation matrix between connected nodes. Conversely, a large maximum independence violation value is expected to correspond to a large correlation matrix value across connected nodes.

As shown in Figure~\ref{fig:natalie}, we consider a star-like arrangement of atoms and two resulting graphs (one unit disk and one disk), generated by setting a global Rydberg blockade radius $r_B$ in the unit disk case and modifying the radii of atoms 0, 4, and 5 in the disk graph case. We fix the global blockade radius to $r_B^G = 7.9055,\mu$m, corresponding to a global drive amplitude of $\Omega = \pi$ rad$/\mu$s. This choice is made using the formula introduced in Section~\ref{sec:lucia} for the global blockade radius, namely $r_B^G = 0.98 \left(C_6 / \Omega\right)^{1/6}$. For atoms 0, 4, and 5, we assign modified blockade radii of $r_{B,c} = 13.0247,\mu$m, such that the corresponding local drive amplitude becomes $\Omega_i = \pi / 20$ rad$/\mu$s. This assignment also follows from the same amplitude–radius relationship established previously. Atoms are arranged with a uniform separation of $0.9r_B$. We analyze three driving protocols: (i) a global pulse applied to the unit disk graph, (ii) a local pulse applied to the disk graph, and (iii) a local drive with randomly shuffled amplitude radii, also applied to the disk graph, to act as a control case. In the global protocol, the amplitude is uniform across all atoms and given by $\Omega=\pi$. In the local protocol, local amplitudes are used to implement local blockade radii; $\Omega_i = \pi/20$ for atoms 0, 4, and 5, and $\Omega_i = \pi$ for all others. In the randomly shuffled protocol, each atom $i$ is independently assigned a local amplitude $\Omega_i = \pi/20$ with probability $3/7$, and $\Omega_i = \pi$ otherwise. 

Figure~\ref{fig:natalie} shows both the maximum independence violation as a function of the scaling parameter $\lambda/\lambda_c$ for the global, local, and randomly shuffled driving protocols, and the corresponding correlation matrices for small and large values of this ratio, in the former two cases. In the correlation matrices, red squares with diagonal crosses denote edges present in the encoded graph. For $\lambda < \lambda_c$ (to the left of the black dotted line), independence is expected to be preserved. This is confirmed for both the global (green diamonds) and local pulses (pink unfilled triangles), where the maximum independence violation remains below 0.2. In contrast, the randomly shuffled protocol (pink filled triangles) exhibits significantly larger maximum independence violations, around 0.8, indicating a breakdown of independence. In this regime, the correlation matrix elements $C_{ij}$ corresponding to edges in both the unit disk and disk graphs remain near 0, consistent with the Rydberg blockade phenomenon. Notably, the disk graph includes three additional edges (0–4, 0–5, and 4–5) relative to the unit disk graph, yet these also show low correlations in the presence of the local drive designed to encode them, confirming that independence is maintained even for these added connections. For larger $\lambda/\lambda_c$ (e.g., $\lambda/\lambda_c = 2$), atoms are spaced farther apart, making double excitations more likely as the conditions for edge encoding are violated. Accordingly, the maximum independence violation approaches 1 across all protocols, and the structure of the correlation matrices changes, with many entries increasing beyond 0.2. 

These observations demonstrate that both the maximum independence violation metric and the correlation matrix can be used to assess how faithfully a given graph has been encoded. In the locally driven case, although we observe pairwise independence as expected, we also sometimes observe it when unexpected, as seen by the presence of low correlation values for some unconnected pairs, for $\lambda/\lambda_c=0.8$. In this case, we have restricted the system to the independent manifold as desired. An ideal embedding would have explored this manifold uniformly, so that all elements of $C_{ij}$ not corresponding to an edge would be of similar values.

\section{Combinatorial Optimization problems}\label{sec:elie}

A prominent motivation for studying Rydberg atom platforms is their suitability for preparing ground states of Hamiltonians that encode combinatorial optimization problems, such as the MIS problem~\cite{Farhi_2001, Das_2008, Albash_2018}. This problem consists of finding the largest subset of vertices in a graph $G$ such that no two are adjacent. In a quantum annealing framework, one aims to transform an initial, easily prepared Hamiltonian into a problem Hamiltonian whose ground state encodes the solution. According to the adiabatic theorem~\cite{Amin_2009}, if the transformation is sufficiently slow, the system remains in its instantaneous ground state. A measurement in the Pauli-$Z$ basis ideally yields the solution to the optimization problem. However, in practice, decoherence, finite evolution times, and nonadiabatic effects often prevent the system from reaching the true ground state. This motivates the study of approximate optimization, which aims to identify near-optimal configurations corresponding to low-energy states with high overlap with the optimal MIS~\cite{Munoz_Bauza_2025}. Approximate solutions are computationally relevant. For large system sizes, preparing independent sets of size $|\mathrm{MIS}| - k$ remains valuable, since the approximation ratio $\frac{|\mathrm{MIS}| - k}{|\mathrm{MIS}|}$
approaches~1 as $|\mathrm{MIS}|$ grows. Moreover, such approximate solutions are experimentally realistic. As discussed in Refs.~\cite{Ebadi_2022, cain2023quantum}, analog quantum hardware naturally produces solutions that are close to the MIS, while preparing the exact optimum remains exponentially difficult. In particular, Ref.~\cite{cain2023quantum} shows that state-of-the-art experiments most frequently yield high-quality approximate solutions rather than true MIS ground states, largely because the dominant difficulty arises from the final step of going from $|\mathrm{MIS}|-1$ to $|\mathrm{MIS}|$.

Given a graph $G=(V, E)$, the Hamiltonian which encodes the MIS problem can be written as:
\begin{equation}\label{eq:MIS}
H_{MIS} = -\sum_{i \in V} n_i + U \sum_{(i, j) \in E} n_i n_j,
\end{equation}
where each vertex $i \in V$ corresponds to a qubit, and where qubits measured in the Rydberg state after annealing represent the vertices in the MIS. The first term energetically rewards the excitation of atoms, and hence the presence of vertices in the set. Choosing $U \gg \Delta(G)$ in the second term, where $\Delta(G)$ is the graph's maximum degree, makes it energetically expensive for neighboring vertices to both occupy the state $\ket{R}$. This ensures low-energy states correspond to MIS solutions.

In Rydberg atom devices~\cite{Ebadi_2022, pichler2018quantumoptimizationmaximumindependent, cazals2025identifyinghardnativeinstances} with Hamiltonian $\mathcal{H}$ as in Eq. ~\eqref{eq:ryd_ham}, the MIS problem can be encoded by setting $\Omega=0$ and choosing atom positions and detuning $\delta$ in order to match Eq.~\eqref{eq:MIS} as closely as possible. This approach has been effectively demonstrated for unit disk graphs with common embedding strategies~\cite{dalyac2024graphalgorithmsneutralatom, Ebadi_2022}. Extensions to more complex graph classes have been achieved either by introducing a large resource overhead through increasing the number of atoms~\cite{Nguyen_2023, lanthaler2024quantumoptimizationgloballydriven} or by using variational methods~\cite{coelho2022efficientprotocolsolvingcombinatorial}. 

In this section, we take a complementary approach: instead of simulating a full annealing protocol, we analyze the ground-state structure of final Hamiltonians that retain finite, locally varying drive amplitudes consistent with disk graph embeddings. Our objective is to determine whether such Hamiltonians, if used as the endpoint of an annealing process, would yield ground states with significant overlap with optimal MIS configurations. This allows us to probe how local drive information, when preserved in the final Hamiltonian, impacts approximate optimization compared to conventional protocols where all drives are suppressed at the end of the evolution.

We approach this by studying the ground-state properties of the Rydberg Hamiltonian after annealing with local, finite amplitudes. In doing so, we aim to determine the most appropriate target Hamiltonian that a quantum annealing algorithm should reach in order to approximately solve MIS problems beyond the unit disk graph class~\cite{breu1998unit, van2006representation}.
It should be noted that the introduction of finite drive as part of the graph encoding shifts the eigenstates of the Hamiltonian away from the Pauli-$Z$ basis, so measurements in this basis become projective. 


We quantify this overlap by comparing the probability of measuring an MIS solution for Hamiltonians constructed with either local or global finite drives. For a given disk graph $G$ with local blockade radii $\{r_i\}_{i=1}^N$ with both local and global, finite drives, we define the problem Hamiltonian $\mathcal{H}(\kappa, \delta_f)$ using local amplitudes $\Omega_i=\kappa C_6/r_i^6$ and a global detuning $\delta_f$. In the global case, the drive amplitude is set to the average value $\Omega=\kappa C_6 /r_{\mathrm{avg}}^6$ where $r_{\mathrm{avg}} = N^{-1} \sum_{i=1}^N r_{i}$
The probability of measuring the MIS is defined as 

\begin{equation}
    P_{MIS}^{dr}(\kappa, \delta_f) = \sum_{\{\textrm{MIS}\}} \lvert \braket{\textrm{MIS} | \psi_0^{dr}(\kappa, \delta_f)} \rvert^2,
\end{equation}

where ${dr} \in\{l, g\}$ indicates the local or global drive, and $\ket{\psi_0^{dr}(\kappa, \delta_f)}$ is the ground state of $\mathcal{H}(\kappa, \delta_f)$ under the respective driving configuration. 

We optimize over the parameters $\kappa, \delta_f$ to find 

\begin{equation}
    P_{MIS}^{dr} = \max_{(\kappa, \delta_f)} P_{MIS}^{dr}(\kappa, \delta_f).
\end{equation}




For the optimized parameters, we first verify that the corresponding ground states preserve the graph-theoretic structure captured by $P_{violation}$ as introduced in Section~\ref{sec:natalie}. This allows us to confirm that the encoded configurations remain consistent with the disk graph topology under local drive. Building on this, we then evaluate approximate optimization performance through the probabilities $P_{\text{MIS-k}}^{dr}$ defined as the probability of obtaining an independent set of size at least $\mathrm{|MIS|}-k$. To quantify the relative improvement provided by local addressing, we introduce the relative sampling enhancement
\begin{equation}
\Delta_k = \frac{2(P_{\text{MIS–}k}^{l} - P_{\text{MIS–}k}^{g})}{P_{\text{MIS–}k}^{l} + P_{\text{MIS–}k}^{g}},
\end{equation}
which compares the near-optimal solution probabilities between local and global drives. We clearly have $\lvert \Delta_k \rvert  \leq 2$ by the triangular inequality. A positive $\Delta_k$ indicates that local driving offers a clear advantage in the approximate optimization regime, and the larger $\Delta_k$, the better local driving is at preparing an approximate MIS.

\begin{figure}[htpb]
\centering
\includegraphics[width=\linewidth]{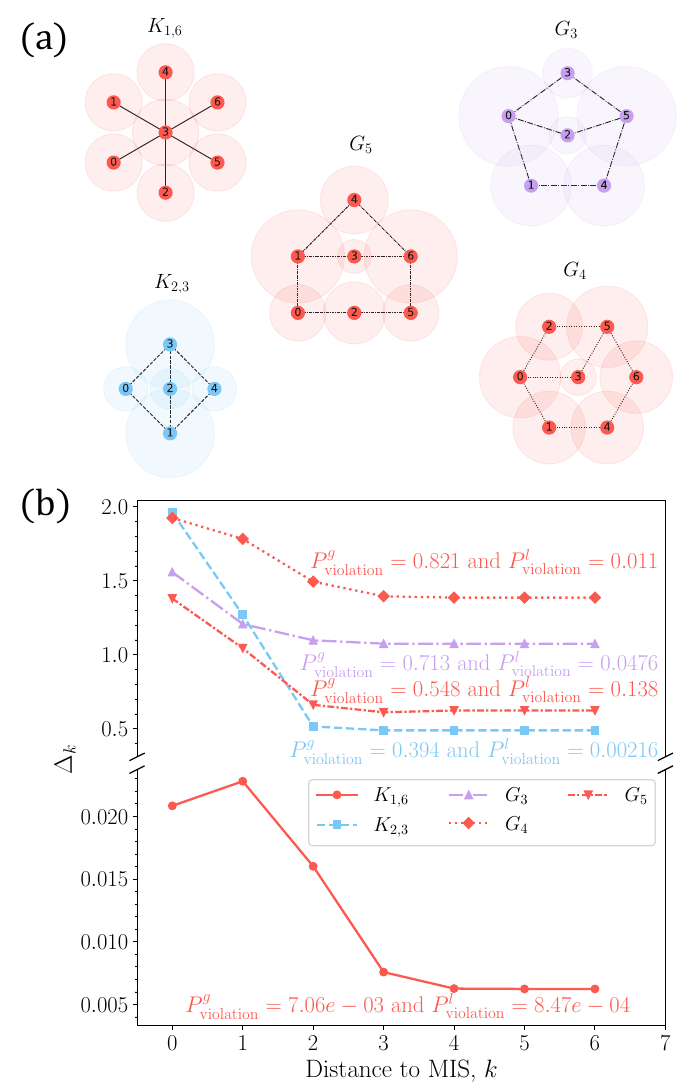}
\caption{
Relative sampling performance of local and global drives. Panel~(a) displays the minimal non-unit-disk graph instances considered. 
For each instance, the parameters $(\kappa, \delta_f)$ are optimized to maximize 
$P_{\mathrm{MIS}}^{\mathrm{dr}}$. 
Panel~(b) shows the relative sampling enhancement $\Delta_k = 
\frac{2\bigl(P_{\mathrm{MIS}-k}^{\,l} - P_{\mathrm{MIS}-k}^{\,g}\bigr)}
     {P_{\mathrm{MIS}-k}^{\,l} + P_{\mathrm{MIS}-k}^{\,g}},$
which compares the probabilities of preparing an independent set of size at least 
$\lvert\mathrm{MIS}\rvert - k$ under local and global driving. 
Annotated values indicate the corresponding violation probabilities 
$P_{\mathrm{violation}}^{g}$ and $P_{\mathrm{violation}}^{l}$, which quantify the probability of leaving the independent-set subspace. Note the difference in scale between the curves for $K_{1, 6}$ and those for the other graphs; this choice highlights the qualitative similarity of the trends while acknowledging their differing ranges. Across these non-unit-disk graphs, local driving systematically reduces 
$P_{\mathrm{violation}}$ and improves the sampling of near-optimal solutions up to $k \approx 2$. 
This illustrates that local addressability is important both for maintaining dynamics within the feasible independent-set subspace and for enhancing approximate MIS preparation.
}
\label{fig:pmis_local_global}
\end{figure}
We apply this analysis to several known minimal non-unit disk graph instances~\cite{non_ud}, shown in Fig.~\ref{fig:pmis_local_global}(a). Figure~\ref{fig:pmis_local_global}(b) reports the relative sampling enhancement $\Delta_k$ together with the corresponding violation probabilities $P_{\mathrm{violation}}$ for these instances. Note the difference in scale between the curves for $K_{1, 6}$ and those for the other graphs; this choice highlights the qualitative similarity of the trends while acknowledging their differing ranges.For each driving configuration, the parameters $(\kappa, \delta_f)$ are optimized to maximize the MIS probability $P_{\mathrm{MIS}}^{\mathrm{dr}}$. Across all instances, we observe that the sampling enhancement is consistently positive and typically large for $k=0$, especially for $K_{2,3}$ and $G_4$. This behavior coincides with a significantly better preparation of independent sets under the local drive compared to the global one: for example, the local drive yields a violation probability approximately $70$ times smaller than the global drive for $G_4$, and roughly $180$ times smaller for $K_{2,3}$. As $k$ increases, the enhancement gradually decreases for $G_4$, while for $K_{2,3}$ it drops sharply at $k=2$, indicating that a global drive becomes competitive only for $\mathrm{MIS}-2$ configurations. For $G_3$ and $G_5$, both curves start around a relative enhancement of $1.5$ at $k=0$ and decrease smoothly with $k$. For $G_3$, the local drive achieves a violation probability about $15$ times smaller than the global scheme. In contrast, $K_{1,6}$ exhibits only a small enhancement for all $k$, with a more pronounced decrease beyond $k=3$. Even in this case, the local drive still maintains a better overlap with the independent-set subspace by approximately one order of magnitude. This behavior can be attributed to the small differences in blockade radii between the central site and the surrounding sites, as seen in Fig.~\ref{fig:pmis_local_global}(b), which reduces the advantage provided by local addressability.

Overall, these results indicate that the global drive often fails to prepare states within the independent-set subspace, whereas local addressing more reliably enforces the constraints and substantially increases the probability of reaching near-optimal MIS configurations across all tested non-unit-disk instances.



These findings highlight that retaining local drive amplitudes as part of the embedding improves the approximate optimization landscape, allowing Rydberg systems to explore a broader class of near-optimal solutions beyond those accessible through global driving and unit disk embeddings.

\section{Conclusion and outlook}


In this work, we revisited the Rydberg blockade phenomenon beyond the conventional model based on a global, fixed blockade radius. Motivated by the need to embed and solve combinatorial problems on more general graph families, we introduced and analyzed the local Rydberg blockade regime, in which atoms experience different drive amplitudes. We characterized this regime using a new quantitative framework, and proposed two metrics; the correlation matrix and the maximum independence violation, to assess how well a physical configuration respects graph-theoretic independence constraints. We proposed and tested a method to solve the MIS problem for this class of graphs using local amplitude and projective measurements, and compared it favorably to a set of control cases of unit-disk graphs with the same average disk radii

Leveraging this framework, we showed that disk graphs, generalizations of unit-disk graphs with local radii, can be deterministically embedded into Rydberg atom arrays through the use of local driving amplitudes. We applied this embedding to the MIS problem in the context of quantum annealing and demonstrated that finite local drives significantly enhance the probability of preparing high-quality approximate solutions on disk-graph instances while staying in the independent set subspace. These results indicate that locally tailored driving schemes can meaningfully extend the computational capabilities of Rydberg platforms. We note that this approach is also directly applicable to unit-disk graphs, recovering previously established results, and can in principle be used to prepare approximate MIS configurations for such graphs within the same embedding framework.

The many body implementation of this embedding remains qualitative, and would benefit from further study in order to quantitatively assess its effectiveness in useful applications. Experimental challenges in annealing to a state with a finite drive may limit the feasibility of the specific MIS algorithm suggested, even for approximation solutions. The extent to which the projective measurement requirement reduces the efficiency also warrants further investigation.

\appendix

\section{Analytical description of the 4 level system for a global simultaneous drive}
\label{app:lucia}
We can diagonalise the Hamiltonian of the two-atom globally driven system with minimal approximation, from which we can obtain the occupation of the doubly-excited state. Although directly differentiating the resulting expression is not achievable, we are able to obtain an expression for the full analytic solution for $P_{RR}$, for any given $r$, $\Omega$ and $\delta$.

Let us rename our variables as $A = -\delta$, $B = \Omega/2$, $C = (C_6/r^6)-2\delta$. Then, we can write $\mathcal{H}$ in matrix form as
\begin{equation}
    \mathcal{H} =
    \begin{pmatrix}
        0 & B & B & 0 \\
        B & A & 0 & B \\
        B & 0 & A & B \\
        0 & B & B & C \\
    \end{pmatrix}.
\end{equation}

To quantify how blockaded a pair of atoms is, we can study the maximum ``leakage'' onto the $\ket{RR}$ state, that is, the peak population of $\ket{RR}$ after evolving the system for a long time. This can be found by calculating $\braket{RR|U(t)|\psi_0}$, where $\ket{\psi_0}$. We can write $\ket{\psi_0}$, $U(t)$ and $\ket{RR}$ in terms of our system's eigenstates and eigenvalues,
\begin{align}
    \ket{\psi_0} &= \sum_i \lambda_i \ket{\psi_i},\quad 
    U(t) = \sum_i e^{-iE_it}\ketbra{\psi_i}{\psi_i}, \\
    \ket{RR} &= \sum_i \beta_i \ket{\psi_i}.
\end{align}
With this, we can extract out the element in $U(t)$ which takes us from our initial state to $\ket{RR}$ as
\begin{align}
    U_{RR} &= \braket{RR|U(t)|\psi_0} \\&= \sum_{i,j,k} e^{-iE_jt}\lambda_i\beta_k \braket{\psi_k|\psi_j}\braket{\psi_j |\psi_i} \\
    &= \sum_{i} e^{-iE_it}\lambda_i\beta_i.
\end{align}
The population of $\ket{RR}$ is then $U_{RR}U^*_{RR}$. Here, it is important to note that all of the entries in $\mathcal{H}$ are real, and hence all $\lambda_i$, $\beta_i$ will be real. With this,
\begin{align}
    (U)_{RR}(U^*)_{RR} &= \left[\sum_{i} e^{-iE_it}\lambda_i\beta_i\right]\left[\sum_{j} e^{-iE_jt}\lambda_j\beta_j\right]^*\\
    &=\sum_i (\lambda_i\beta_i)^2 + 2\sum_{i>j}(\lambda_i\beta_i)(\lambda_j\beta_j)\cdot\text{cos}(\Delta E_{i,j}t).\label{eq:dynamics}
\end{align}

Now, we can use the property that, for two periodic functions $f(x)$ and $g(x)$ with periods $a$ and $b$, respectively, the period of $f(x)+g(x)$ will be the least common multiple between $a$ and $b$. This means that $\sum_{i>j}\text{cos}(\Delta E_{i,j}t)$ will (at least, approximately) saturate at 1 for a long enough time. Then, we can reduce the above equation to 
\begin{equation}
    (U)_{RR}(U^*)_{RR}^{\text{max}} =\sum_i (\lambda_i\beta_i)^2 + 2\sum_{i>j}|(\lambda_i\beta_i)(\lambda_j\beta_j)|. \label{eq:upper_bound}
\end{equation}


The eigenvectors of our Hamiltonian can be written as follows,
\begin{align}
    & v_1 = (0,-1,1,0)\\
    & v_i = \left(\dfrac{E_i-C}{E_i}, \dfrac{E_i-C}{2B}, \dfrac{E_i-C}{2B}, 1 \right) \quad\text{for } i=2,3,4,
\end{align}
so, for example, if we start with our system in the double ground state, $\ket{gg}$, we will have
\begin{align}
    (UU^*)_{\ketbra{RR}{gg}}^{\text{max}} &= \sum_{i=2} \left(\dfrac{E_i - C}{E_i} \cdot \dfrac{1}{L_i^2} \right)^2 \\
    &\quad + 2\sum_{i>j=2} \left| \left(\dfrac{E_i - C}{E_i} \cdot \dfrac{1}{L_i^2} \right) \left(\dfrac{E_j - C}{E_j} \cdot \dfrac{1}{L_j^2} \right) \right|.
\end{align}
where $L_i$ is the length of vector $i$, i.e.,
\begin{equation}
    L_i = \sqrt{\left(\dfrac{E_i-C}{E_i}\right)^2 +2\cdot\left( \dfrac{E_i-C}{2B}\right)^2+ 1 }\;.
\end{equation}
The eigenvalues can be written as follows,
\begin{align}
    & E_2 = \dfrac{1}{3}\left(A+C-2\sqrt{P}\cdot\text{cos}\left(\dfrac{\text{tan}^{-1}(Q/M)}{3}\right)\right) \\
    & E_3 = \dfrac{1}{3}\Bigg(A+C+\sqrt{P}\cdot\text{cos}\left(\dfrac{\text{tan}^{-1}(Q/M)}{3}\right) \\
    & \quad -\sqrt{3P}\cdot\text{sin}\left(\dfrac{\text{tan}^{-1}(Q/M)}{3}\right)\Bigg) \\
    & E_4 = \dfrac{1}{3}\Bigg(A+C+\sqrt{P}\cdot\text{cos}\left(\dfrac{\text{tan}^{-1}(Q/M)}{3}\right) \\
    & \quad +\sqrt{3P}\cdot\text{sin}\left(\dfrac{\text{tan}^{-1}(Q/M)}{3}\right)\Bigg),
\end{align}

where,
\begin{align}
    & P = -3AC +12B^2 + (A+C)^2\\
    & M = 27B^2C+\dfrac{9}{2}(A+C)(AC-4B^2)-(A+C)^3\\
    & Q = \sqrt{P^3-M^2}\;.
\end{align}

We observe that, for $\delta=0$, for all three systems, $P_{RR}$ is symmetric around the value 0.5, i.e. around $r_B$. Motivated by this, and with the goal of writing a more simple expression to well approximate $P_{RR}^{\text{G}}$ (i.e. $P_{RR}$ for the global system) we translate the sequential (global) curve by $r_B$($r_B^{\text{G}}$) and scale by a factor $\Delta^\pi$($\Delta^{\text{G}}$, defined that the gradient at $r=r_B$($r_B^{\text{G}})$ is 1. We determine the value of  $\Delta^\pi$ analytically and that of $\Delta^\text{G}$ numerically using the expression for $P_{RR}^\text{G}$. The resulting gradients are
\begin{equation}
    \begin{split}
        \Delta^\pi &= \frac{\partial P_{RR}^\pi}{\partial r} \bigg\rvert_{r=r_B^\pi} = \frac{3}{r_B^\pi}\\
        \Delta^{\text{G}} &= \frac{\partial P_{RR}^{\text{G}}}{\partial r} \bigg\rvert_{r=r_B^G} = \frac{3.864}{r_B^\pi},
    \end{split}
\end{equation}
and can be used to map the global curve onto the sequential one by mapping the atom separation $r$ as follows
\begin{equation}
    r\rightarrow \frac{\Delta^{\text{G}}}{\Delta^\pi}(r-r_B^{\text{G}})+r_B^{\pi}.
\end{equation}
With this we can write a much simplified expression for $P_{RR}^G$,
\begin{equation}\label{eq:prr_globpulse}
    P_{RR}^{\text{G}} = \dfrac{1}{1+\left[\dfrac{1}{\Omega}\cdot\left(\dfrac{C_6}{(1.29r-0.26r_B)^6}\right)\right]^2},
\end{equation}

\section{Locally driven blockade radius} 
\label{app:local_rydberg}

\begin{figure}[htbp]
        \centering
        \includegraphics[width=\linewidth]{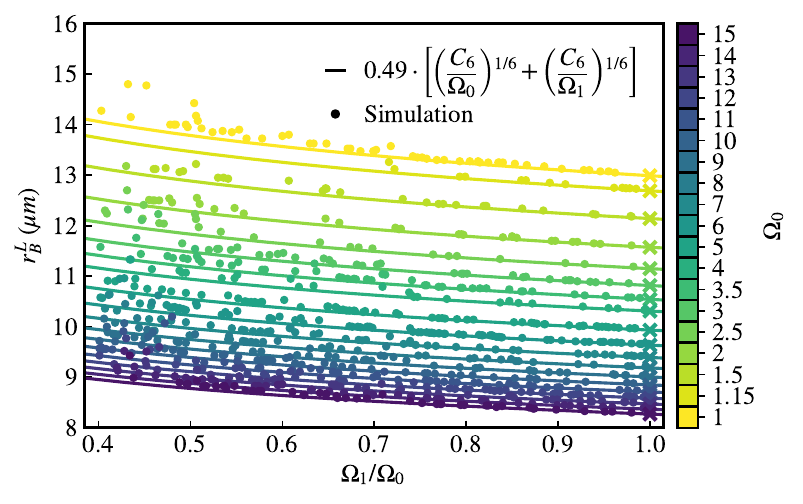}
        \caption{
        Blockade radius for systems with local simultaneous drive with different $\Omega_0$ and $\Omega_1$, obtained by simulation (circular points). Pulse durations for each simulation were chosen differently. Crosses mark data points where $\Omega_0=\Omega_1$ and solid lines indicate the value of $r_B^\text{L}$ found using Eq.~\eqref{eq:rb_l}.}
        \label{fig:rb_local}
\end{figure}

For locally driven systems we restrict ourselves for simplicity to the zero detuning case with Hamiltonian
\begin{equation}\label{eq:ryd_ham_local}
    \mathcal{H} = \dfrac{1}{2}\sum_i \Omega_i \sigma_i^x + \sum_{i\neq j} \dfrac{C_6}{r_{ij}^6}n_in_j,
\end{equation}
which, even for a 2 atom system is nontrivial to diagonalise. As a result we cannot solve the Schr\"odinger equation to find an analytic expression for the maximum occupation of the doubly-excited state as we did for the sequential and global systems. Instead, we use the same observation of near-symmetry as when building a simplified expression for global simultaneous driving system to find $P_{RR}^\text{L}$ (i.e. $P_{RR}$ for the local system) by rescaling $r$ and using the simple expression for $P_{RR}^\pi$. This requires an expression for $r_B^\text{L}$ ($r_B$ for the local system) and $\Delta^\text{L}$ (the gradient of $P_{RR}^\text{L}$ around $r_B^\text{L}$) to implement
\begin{equation}\label{eq:rescaling_local}
    r\rightarrow (\Delta^{\text{L}}/\Delta^\pi)\cdot[r-r_B^{\text{L}}]+r_B^{\pi}.
\end{equation}
The blockade radius $r_B^\text{L}$ is found by choosing a symmetric expression across both atoms which obeys the value found using the global system) for $\Omega_0=\Omega_1$ to match,
\begin{equation}\label{eq:rb_l}
    r_B^\text{L} = \frac{1}{2}\left(r_{B, 0}^\text{G} + r_{B, 1}^\text{G}\right),
\end{equation}
where $r_{B, i}^\text{G}$ is the blockade radius one would expect from a globally driven system with amplitude $\Omega_i$. We compare this to values found from simulated data for different $\Omega_0$, $\Omega_1$ values, in Figure \ref{fig:rb_local}, using simulations sampling from a restricted set of distances around the expected value of $r_B^\text{L}$, such that a high resolution in $r$ can be achieved. For all simulations in this section we evolve for 15$\mu$s. Using simulation to obtain $P_{RR}$ values through can result in lower values being obtained than the true maximum values, if the pulse duration used is not sufficient to reach the maximum. This leads to larger inferred values of the blockade radius than accurate. As such, we expect that Eq.~\eqref{eq:rb_l} to set the lower bound for the simulation-inferred values of $r_B^\text{L}$, as seen in Figure \ref{fig:rb_local}.  Data points with lower $\Omega_1/\Omega_0$ were rarely satisfied the criteria used to ensure good approximation, so we restrict the ratio range to $\Omega_1/\Omega_0 \in [0.4,1]$.

It is clear from Eq.~\eqref{eq:rb_l} that we could instead use the global expression for the pairwise blockade radius and instead consider an effective amplitude of the local system to be the generalized mean of the two amplitudes with exponent $-1/6$, 
$\Omega_\text{eff} = \left(\frac{1}{2}[\Omega_0^{-1/6}+\Omega_1^{-1/6}]\right)^{-6}$.

\begin{figure}[htb!]
        \centering
        \includegraphics[width=\linewidth]{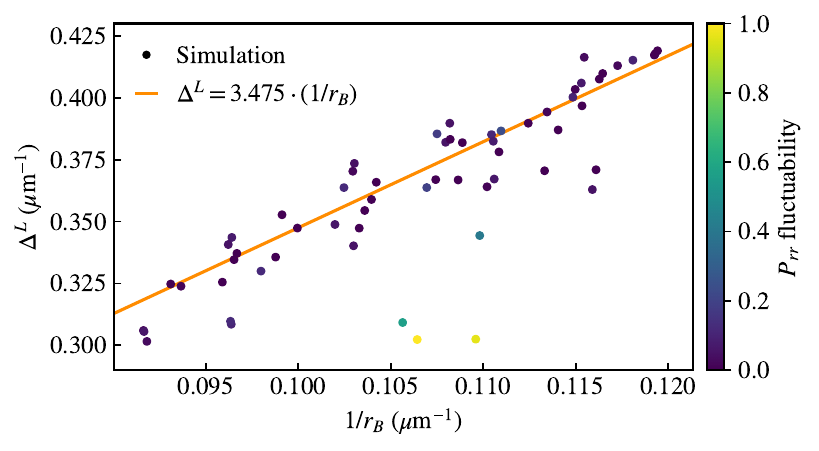}
        \caption{
        Gradient for systems with local simultaneous drive obtained by minimizing the distance from Eq.~\eqref{eq:prr_pipulse} to simulated data, shifted and rescaled according to $r \to (\Delta^{\text{L}}/\Delta^\pi)\cdot[r-r_B^{\text{L}}]+r_B^{\pi}$. Points of higher fluctuability are shown lighter.
        }
        \label{fig:grad_local}
\end{figure}

We repeat the process of finding a gradient $\Delta^\text{L}$ using simulation results by applying the rescaling described in Eq.~\eqref{eq:rescaling_local} to simulated data over a range of distances ($5\,{\text -}\,24.75\mu$m) and finding value of $\Delta^\text{L}$ which minimises the least-squares difference between $P_{RR}^\pi$ and the rescaled $P_{RR}^\text{L}$. This was repeated for 61 combinations of $\Omega_0$ and $\Omega_1$ and the resulting values of $\Delta^\text{L}$ are shown as a function of  $1/r_B(\Omega=\Omega_\text{eff})$ in Figure \ref{fig:grad_local}. We then fit to these data points to find the relation
\begin{equation}\label{eq:grad_l}
    \Delta^{\text{L}} = \frac{\partial P_{RR}^\text{L}}{\partial r} \bigg\rvert_{r=r_B^\text{L}} = \frac{3.475}{r_B^\pi(\Omega=\Omega_\text{eff})}.
\end{equation}
For a small number of the simulations we observe large fluctuations of $P_{RR}^\text{L}$ at random values of $r$, despite the expectation that the function is monotonic. We attribute a higher error to simulations with larger fluctuations and these errors are then taken into account in the fitting procedure in Figure \ref{fig:grad_local}, where data points with larger fluctuations contributing less to the fit. The `fluctuability' of each simulation was calculated by tracking the changes in $P_{RR}^\text{L}$ within non-monotonic regions of the data. This fluctuability was considered when fitting the data points to a linear function, where the fit is found by minimizing $\sum_i d_i\cdot(1-F_i)$, where $d_i$ is the distance between data point $i$ and the fit, and $F_i$ is the fluctuability of point $i$.

\begin{figure}[htb!]
        \centering
        \includegraphics[width=\linewidth]{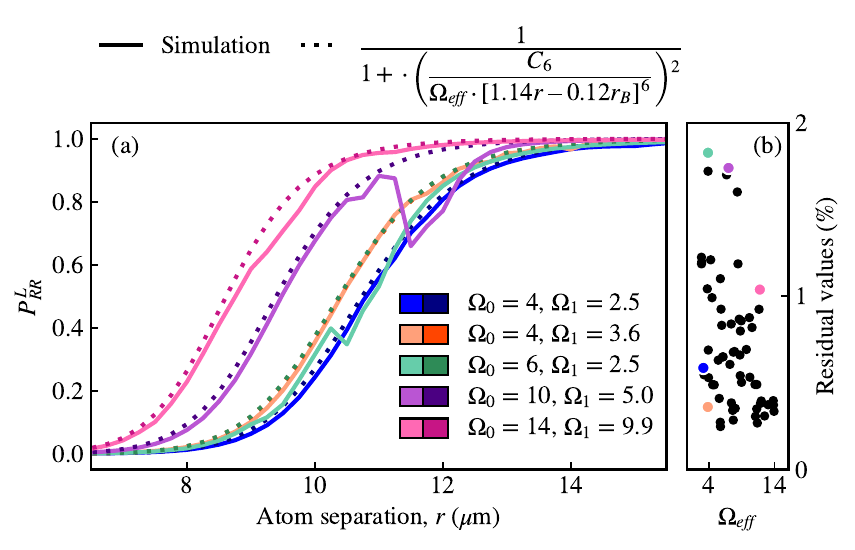}
        \caption{
        (a) Maximum occupation of the doubly-excited state for five simulations of local simultaneous systems and Eq. ~\eqref{eq:prr_localpulse}. (b) For 61 of such simulations, the residual values are shown, with the 5 which correspond to the data in (a) depicted in color.
        }
        \label{fig:prr_local}
\end{figure}

This enables us to write
\begin{equation}\label{eq:prr_localpulse}
    P_{RR}^{\text{L}} = \dfrac{1}{1+\left[\dfrac{1}{\Omega_\text{eff}}\cdot\left(\dfrac{C_6}{(1.18r-0.16r_B)^6}\right)\right]^2},
\end{equation}
which is plotted in Figure \ref{fig:prr_local} (a) for five of the studied simulations and shows good agreement with simulation. We again note that we expect simulation to underestimate $P_{RR}$. The residual distances between Eq.~\eqref{eq:prr_localpulse} and all 61 simulations are shown in Figure \ref{fig:prr_local} (b), with none exceeding 2\%.

\bibliographystyle{apsrev4-1}
\bibliography{refs}

\end{document}